\documentclass[aip,graphicx,reprint,amsmath,amssymb]{revtex4-1}
\usepackage{graphicx}
\usepackage{dcolumn}
\usepackage{bm}
\usepackage[draft]{changes}

\begin{document}

\preprint{}

\title{The Hall Term and Anomalous Resistivity Effects in Neon Gas-Puff Z-Pinches}

\author{A. Rososhek}
    \email{ar877@cornell.edu}
\author{C. E. Seyler}
\author{E. S. Lavine}
\author{D. A. Hammer}
\affiliation{Laboratory of Plasma Studies, Cornell University Ithaca, New York 14853, USA
}%


\begin{abstract}
In this paper, we compare experimental and numerical simulation results to benchmark the PERSEUS code against gas-puff $Z$-pinch implosions on COBRA.  We then use the code to investigate the structure of the plasma sheath. To this end, we study the morphology of the implosion, focusing on non-magnetohydrodynamical (MHD) effects such as electron drifts governed by the Hall term within the growing magneto-Rayleigh-Taylor instability (MRTI). The spatial wavelength of MRTI is better reproduced when both the Hall term and an anomalous resistivity driven by the electron drift are included. Additionally, cathode-anode gap polarity effects are more accurately captured when the Hall term is turned on. The plasma sheath structure, which includes both the accelerating piston driven by the magnetic pressure and the shockwave ahead of it, matches interferometric measurements in width only when a current-driven anomalous resistivity model is used. This anomalous resistivity is assumed to be driven by the lower-hybrid-drift instability, which generates small-scale turbulence with typical wavelengths $<30\mu m$.
\end{abstract}

\maketitle

\section{\label{sec:intro}Introduction\protect
}
\par High-energy-density (HED) plasmas, such as those formed in fast $Z$-pinches, exhibit extreme conditions where classical resistivity models fail to accurately describe the physics.\cite{Ryutov2015,Hares_1985} Traditional magnetohydrodynamic (MHD) models rely on Spitzer resistivity, which assumes a collisional plasma with a relatively simple temperature-dependent ($T$) resistivity ($\eta$) $\eta\propto T^{-3/2}$.\cite{Spitzer_res} For example, in gas-puff $Z$-pinches (GZP), where plasma is being heated, compressed, exhibits turbulence, and is subject to Hall physics, the resistivity is almost certainly not well-described by the Spitzer model and can be considered anomalous. While direct experimental evidence is sparse and the primary driving mechanism remains unidentified, micro-instabilities such as the Lower-Hybrid Drift Instability (LHDI) are likely to be involved.\cite{Sander_iaw} Additionally, observations of the lower hybrid microinstabilities have been reported to be present in fiber $Z$-pinches affecting its performance.\cite{Chittenden_95}
\par The electrical conductivity of HED plasmas is governed not only by electron-ion collisions but also by collective plasma effects, wave-particle interactions, and kinetic instabilities.\cite{Fisch21,Montgomery_LPI} Experimental studies on GZP show deviations from classical resistivity predictions, with measured resistivity often being orders of magnitude higher than expected.\cite{Giuliani15} This discrepancy likely stems from instabilities such as MRTI and LHDI, electron heating, and non-Maxwellian distribution functions. Hall physics introduces additional complexity, particularly in low-density power-feed plasmas surrounding the main imploding plasma column.\cite{Woolstrum2022} The Hall physics become important when both the Hall parameter $\beta_H\equiv\omega_{ce}\tau_e=\frac{eB}{m_e\nu_{ei}}$ ($\omega_{ce}$ is the electron cyclotron frequency, $\tau_e$ is the electron relaxation time, $e$ is the electron charge, $B$ is the magnetic field, $m_e$ is the electron mass, $\nu_{ei}$ is the electron-ion collision frequency) and the ion inertial length, $c/\omega_{pi}$ ($c$ is the speed of light and $\omega_{pi}$ is the ion plasma frequency), are comparable to relevant scale lengths. Under such conditions, Hall MHD effects significantly modify current flow and magnetic field transport, leading to nontrivial resistive dissipation.\cite{Rahman_12} These effects cannot be captured by standard resistive MHD models, necessitating advanced two-fluid or hybrid kinetic-fluid simulations. To facilitate such modeling efforts and more accurately incorporate Hall and other non-ideal plasma physics, the extended MHD code PERSEUS was developed.\cite{Seyler_Perseus} Additionally, as noted in Ref. [\onlinecite{McBride_10}], the two-fluid extended MHD model can drive asymmetrical axial flows that produce anode-cathode asymmetry, the so-called polarity effects, observed in a tungsten wire-array z-pinch experiment.
\par Recent research on the GZP shows that the plasma sheath, which includes the current-carrying layer (piston), a shockwave, and a region in between, is turbulent.\cite{Kroupp_turbo,Sander_iaw,Rocco_epw,Ros_rand} Moreover, evidence from interferometric measurements, for example, found in Ref. [\onlinecite{Sander_impldyn}], shows the inconsistency between the collisions-driven Spitzer resistivity model used to estimate the skin depth width of the plasma sheath and the observed value. For example, the skin depth scales as $\propto\sqrt{\eta}\propto T^{-3/4}$, where $\eta$ is the plasma resistivity and $T$ is plasma temperature. According to the classical Spitzer prediction, the plasma sheath width should decrease with increasing temperature, which is not the observed behavior on COBRA.\cite{Angel2024, Sander_iaw} A new study based on polarization Zeeman spectroscopy found that the current-carrying layer within the plasma sheath early into a run-in stage ($\approx21~mm$ plasma radius) is $\lesssim 1~mm$ wide.\cite{Angel2024} Moreover, by investigating the stopping lengths of the upstream ions within the sheath, it was shown that this interaction is collisionless, which indicates a more intricate ion kinetic energy dissipation mechanism.\cite{Sander_iaw} These observations indicate that the conditions within the plasma sheath structure are not well understood, and the existing resistivity model requires revisiting.
\par Polarity effects affecting the electrical explosion of single wires and wire arrays in vacuum and different media have been studied for decades; however, the main focus was to determine the energy deposition effects, resistivity, and changes in the ablation rate.\cite{Russkikh2006,Bland_05} In GZP experiments, these effects are largely unexplored and the research is focused on the stagnation stage.\cite{Imasaka_01,Conti_20}

\par In this paper, we employ the numerical code PERSEUS\cite{Seyler_Perseus} to study different morphology-related effects in GZP while also using new and prior experimental evidence to link these effects to Hall physics and current-driven anomalous resistivity. We take advantage of the PERSEUS code's being massively parallel and run it at the high performance computing center in Pittsburgh, the Bridges-2 regular memory facility.\cite{bridges2} The main goal of this paper is to present the importance of including the Hall term and using anomalous resistivity to simulate the gas-puff z-pinch experiments. While some comparison with the experimentla data will be provided, at this stage we cannot execute a closely tailored simulation run to the experimental campaign, which we hope to do in the future.
\par This paper is structured as follows: we begin with a detailed description of the simulation framework, including the model, boundary conditions, and initial conditions used to study GZP. Next, we provide a brief overview of the pulsed power facility, outlining its key parameters relevant to the experiments and diagnostics used for comparison. We then present and analyze the numerical results, comparing them with experimental data to assess model validity and identify key physical mechanisms. Finally, we summarize our findings and discuss their implications for future studies in the conclusion section.

\section{\label{sec:expsetup}Experimental setup\protect
}
\subsection{\label{subsec:exp}Gas-Puff Z-pinch on COBRA}
\begin{figure}
    \centering
    \includegraphics[width=0.5\linewidth]{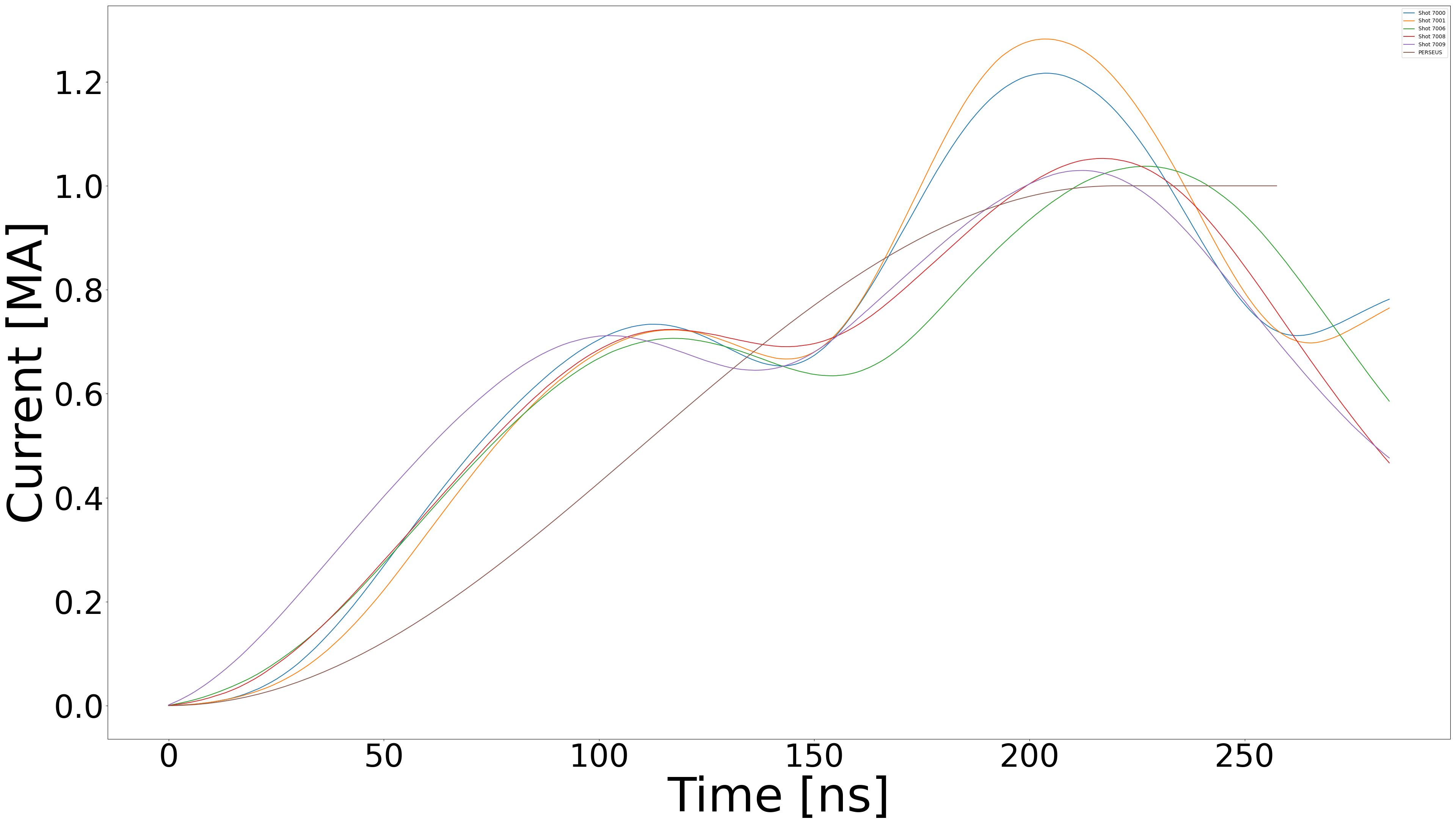}
    \caption{The measured COBRA current during one experimental compaign along with the current form used in the PERSEUS code run.}
    \label{fig:current}
\end{figure}

The COBRA pulsed power generator\cite{Cobra} is a low-impedance, $0.5~\Omega$, driver that includes two Marx generators, each consisting of 16 capacitors of $1.35~\mu F$ charged to $70~\text{kV}$, yielding approximately $105~\text{kJ}$ of stored energy. This machine is designed to drive a $1~MA$ current with a variable $100-220$ ns rise-time through a low-inductance, typically about $10~nH$, the load in this case is a custom gas-puff nozzle designed in cooperation with the Weizmann Institute. Throughout the experimental campaign, the three independently pressurized nozzles were filled with Ne gas with pressures adjusted for a machine over-massed load, i.e. when the load mass exceeds what the available current can compress efficiently, $\approx4.4~\mu g/mm$, that is $\approx25~mm$ long. The load mass profile was measured using planar laser-induced fluorescence (PLIF) and reported in Ref. [\onlinecite{deGrouchy_plif}]. Typical experimental results from operating COBRA in a machine-overmassed regime were used as the basis for the numerical code runs. Typical oscillograms of various overmassed Ne shots of the same experimental campaign are shown in Fig. \ref{fig:current} along with the current used in PERSEUS. A more detailed information together with the experimental scheme can be found in Refs. \onlinecite{Sander_iaw, Sander_impldyn}.

\subsection{\label{subsec:sim}PERSEUS simulation}
The PERSEUS code runs for this paper use a 3D implementation of Extended magneto-hydrodynamical (XMHD) model, which solves the XMHD equations outlined in Refs. [\onlinecite{Woolstrum2022},\onlinecite{Seyler_Perseus}]. This code includes an additional physics package beyond ideal MHD, i.e. the generalized Ohm’s law, incorporating Hall terms, resistive effects, and anomalous resistivity based on Refs. [\onlinecite{Davidson_1977, Dvdsn_Gldd}]. From Ref. [\onlinecite{Dvdsn_Gldd}], one can deduce a model for the LHDI-driven resistivity, as shown in Eq. \ref{eq:anom_res2}. To estimate the fluctuating electric field energy, $|\delta E|^2$, the authors suggest that it may saturate due to ion trapping, leading to the following scaling based on a thermodynamic bound: $\frac{\epsilon_0|\delta E|^2}{2}\sim \sqrt{\frac{\pi}{2}}\frac{m_e}{m_i}n_eT_e$, where $\epsilon_0$ is the vacuum permittivity, $m_e$ and $m_i$ are the electron and ion masses, respectively, $n_e$ is the electron density and $T_e$ is the electron temperature. This is an interesting result as it shows that the LHDI-driven resistivity would increase with temperature as opposed to the classical Spitzer resistivity model. 
\par From the results of Davidson and Gladd,\cite{Dvdsn_Gldd} the resistivity model is taken to be $\eta=\eta_s+\eta_*$, where $\eta_s$ is the usual Sptizer resistivity and the anomalous resistivity $\eta_*$ is taken to be in SI units.

\begin{equation}
    \eta_*=\frac{m_e\nu_{\rm eff}}{n_ee^2}\approx{\rm min}\left[\sqrt{\frac{\pi}{2}}\frac{m_e}{m_i}\frac{\alpha(v_{de}/v_i)^2}{1+(v_{de}/v_i)^2}\frac{1}{\epsilon_0\Omega_e},\frac{B}{n_ee}\right]
    \label{eq:anom_res2}
\end{equation}
where $\nu_{\rm eff}$ is the effective collision frequency due to the instability, $\alpha$ is an adjustable parameter related to the saturation amplitude of the instability and is assumed to be of order unity, $v_{de}$ is the perpendicularly resolved electron drift speed determined from the current given by $v_{de}=|{\bf u}-{\bf J}/n_ee|$, where ${\bf u}$ is the flow speed and $\bf J$ is the current density, $v_i$ is the ion thermal speed, and $\Omega_e$ is the electron cyclotron frequency, and $B$ is the magnetic field. The term $B/n_ee$ is taken to be the maximum value the resistivity can attain based on the condition $\nu_{\rm eff}<\Omega_e$, otherwise the electrons would be unmagnetized and the LHDI would not occur. Note that the dependence of $\eta_*$ on the current scales as its square for drift speeds much lower than the ion thermal speed. This is consistent with the quasi-linear transport analysis of Davidson and Gladd for the low-drift speed case in which the LHDI growth rate is proportional to $v_{de}^2$.

\par It is important to point out that Eq. \ref{eq:anom_res2} is not exactly the same as Eq. (40) in the paper by Davidson and Gladd. The main differences are: the SI units used here; the dependence on the electron drift speed, which is approximated by the low-drift speed formula $(v_{de}/v_i)^2/(1+(v_{de}/v_i)^2)$; the saturation amplitude has the additional dimensionless phenomenological parameter $\alpha$; and the term $B/n_ee$ is taken to be the maximum value the resistivity can attain based on the condition $\nu_{\rm eff}<\Omega_e$, otherwise the electrons would be unmagnetized and the LHDI would not occur. We believe that Eq. (1) is a reasonable phenomenological model representing the effect of the LHDI on the resistivity. Part of the intent of this paper is to a certain extent validate Eq. (1) through comparisons of Hall-MHD simulations using PERSEUS with experiments of GPZ-pinches.

\par In the current realization of PERSEUS code the plasma self-emitted radiation is neglected. The simulation data used for the current paper taken before the stagnation stage onset where the plasma density and temperature across the plasma sheath is relatively low; hence, we assume that the amount of radiated energy is negligible and has limited effect on the pre-stagnation phase.$^{00}$ In Ref. [4], Fig. 3c a generic PCD signal is shown for Ne overmassed implosion, where almost no X-ray observed before stagnation, which supports the assumption of negligible radiated energy output prior to the stagnation stage onset.

\par Boundary conditions in PERSEUS runs were adjusted to reflect the GZP experimental setup on COBRA\cite{Sander_impldyn}, with the bottom and side boundaries modeled as conducting within the cathode radius and supporting an azimuthal magnetic field outside, including along the sides. The top boundary was taken to be conducting. Due to the time limitations at the Bridges-2 facility, we have run PERSEUS in two different configurations. The configuration \#1 was designed to enable more runs for less time with the main goal to test the model-specific features. In this configuration, the simulation volume was $60\times60\times12.5~mm^3$ on a $1000\times1000\times200$ grid. The configuration \#2 was designed to match the experimental run on COBRA as close as possible. The domain size is then realistic, $60~mm$-by-$60~mm$ in $x$, $y$, and ~$24.2~mm$ in $z$-axis respectively, which is about equal to the real gas-puff while the boundary-related effects are limited. To maximize spatial resolution via an MPI-enabled numerical scheme we run these simulations on 16 regular memory nodes, 125 cores each, enabling the simulation volume of $960\times960\times400=3.6864\times10^8$ cells. Each voxel has a size of about $0.0625~mm\times0.0625~mm\times0.061~mm\approx0.24\times10^{-3}~mm^3$. The timesteps in this simulation are different, i.e. for $t<195~ns$ it equals to $3~ns$ and for $t>195~ns$ it is roughly $0.5~ns$.

\section{\label{sec:res_diss}Results and Discussion\protect
}
To demonstrate the importance of including the Hall term-related physics and anomalous resistivity in GZP conditions, we used Mach-Zehnder interferometry, extreme ultraviolet (XUV), and shadowgraphy data from Ref. [\onlinecite{Sander_iaw}]. Using these data we studied the morphology of the GZP, which involves comparison of the spatial frequency and shape of the MRTI bubbles and spikes at different time-steps during implosion, the polarity effect, and the measured width of the plasma sheath. Note, that these comparisons are qualitative to support the main argument of this paper about the importance of Hall term-related physics and anomalous resistivity. The Hall term in the generalized Ohm's law (GOL), outlined in Ref. [\onlinecite{Seyler_Perseus}], is equal to $-\frac{1}{ne}\mathbf{J}\times\mathbf{B}$. It is clear that Hall term related effects should be expected outside of the plasma sheath - at the piston-magnetic field interface and further out inside the MRTI's, and generally in the low-density regions of the pinch. The Hall term generates an electric field directed radially outward that is stronger near the cathode, resulting in stronger pinching. That electric field also drives a radial component in the electron velocity that, in turn interacting with the magnetic field, leads to an $\mathbf{E}\times\mathbf{B}$ drift in $+\hat{\mathbf{z}}$ direction. Thus, by including a Hall term in the XMHD model one can expect to capture this effect.
\par
\begin{figure}
    \begin{tabular}{cc}
        \centering
        \includegraphics[width=0.5\linewidth]{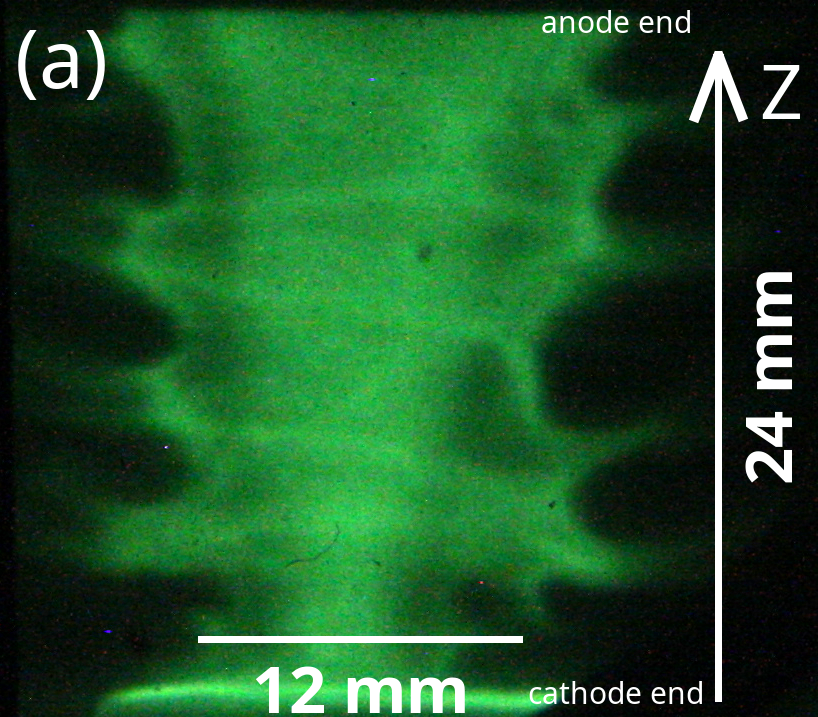}
        \includegraphics[width=0.5\linewidth]{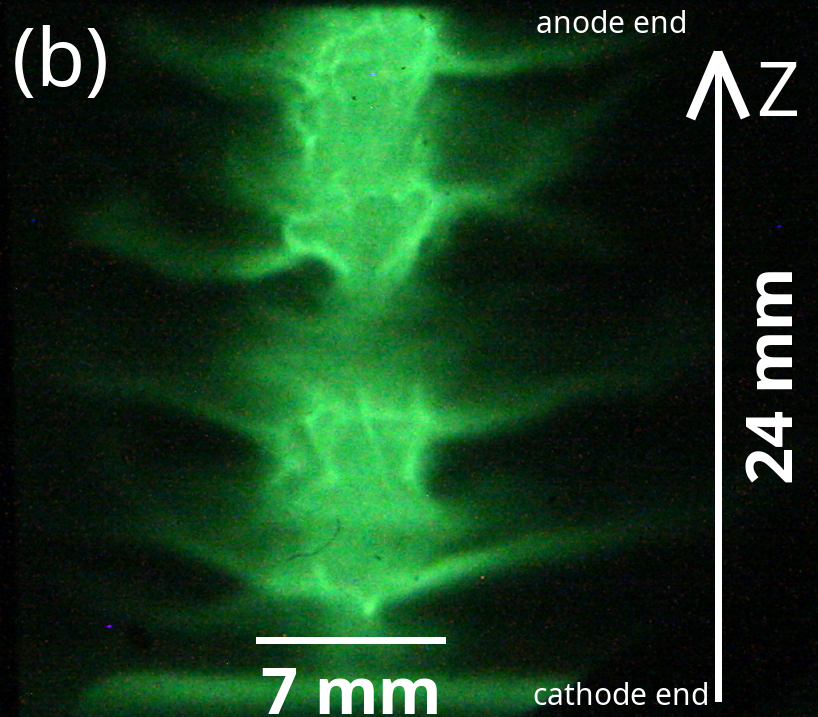}
    \end{tabular}
    \caption{The typical XUV images representing key morphology features of the Ne GZP experiment on COBRA}.
    \label{fig:exp_data}
\end{figure}

 Previous research on the shock-piston layer and the MRTI in GZPs,\cite{Grouchy_shocks,Sander_impldyn} found that the MRTI spatial frequency is almost constant during the run-in stage for Ne, Ar, and Kr gases indicating a generally applied dynamics of the MRTI development, including that of the MRTI bubble interior. In Fig. \ref{fig:exp_data}, we present the XUV images that represent additional effects, such as the generally observed directionality in $
+\hat{\mathbf{z}}$ direction, i.e. from the cathode to anode, of the MRTI bubbles outside edges. Similar feature can be found in Ref. [\onlinecite{Bland_05}], where the authors studied the effect of the externally applied radial electric field on the ablation rate during wire array z-pinch experiments. In Fig. 5(b) of that paper, the authors show XUV images of the implosion of a wire array where the MRTI bubbles seem to have their outside edges' direction dependent on the radial electric field direction. 
\par In Fig. \ref{fig:exp_data} we show the routinely observed gas-puff z-pinch morphology measured experimentally on COBRA-driven implosions. Here, we would like to note the upward direction of the MRTI bubbles and early pinching at the cathode. As mentioned before, this configuration may be created by the radial electric field that will drive the electron velocity in the outward direction. Then this drift velocity in $+\hat{\mathbf{r}}$ direction interacting with the azimuthal magnetic field will produce an $\mathbf{E}\times\mathbf{B}$ drift in the $+\hat{\mathbf{z}}$ direction generating the directionality. 
\begin{figure}
    \begin{tabular}{cc}
        \includegraphics[width=0.5\linewidth]{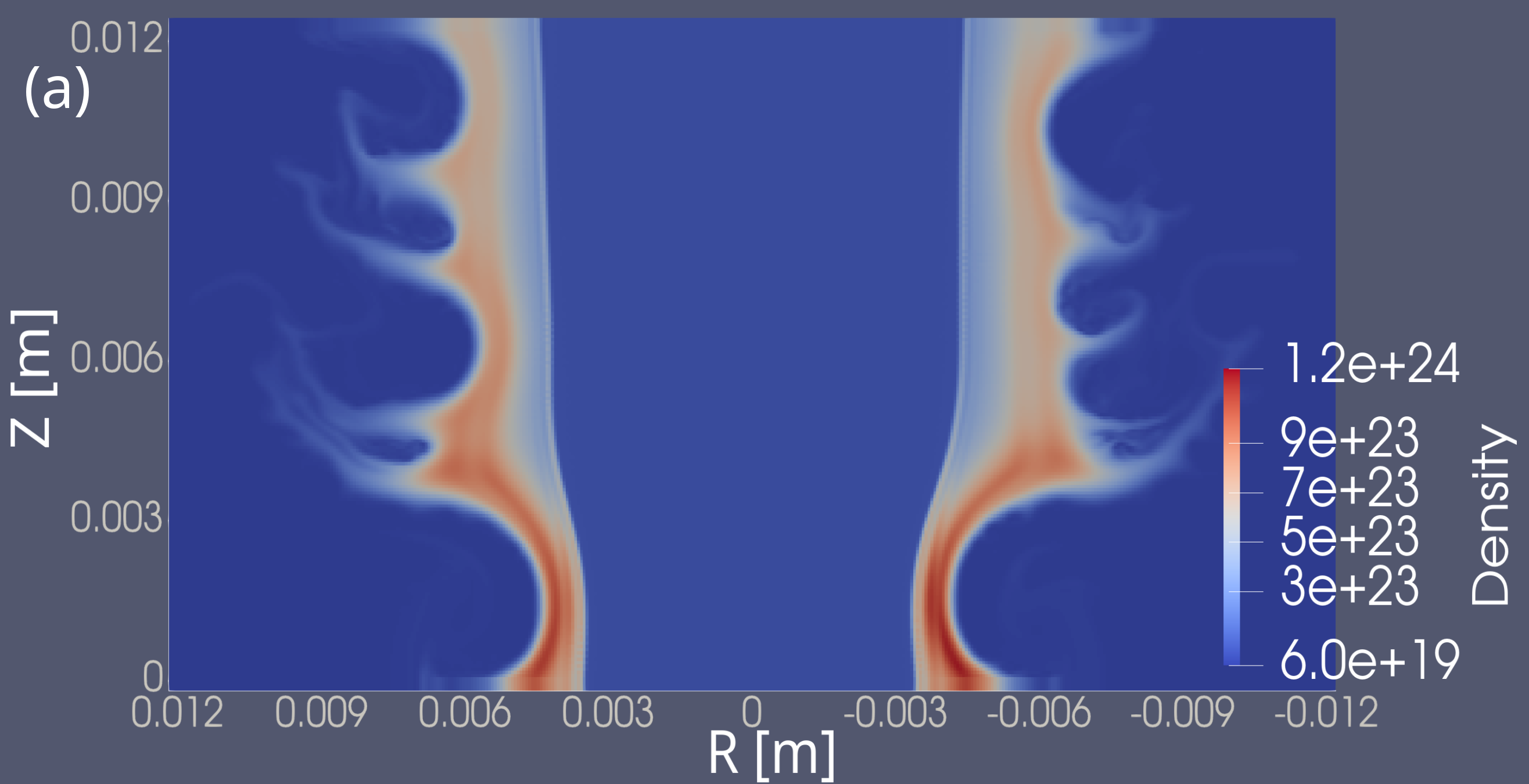}
        \includegraphics[width=0.5\linewidth]{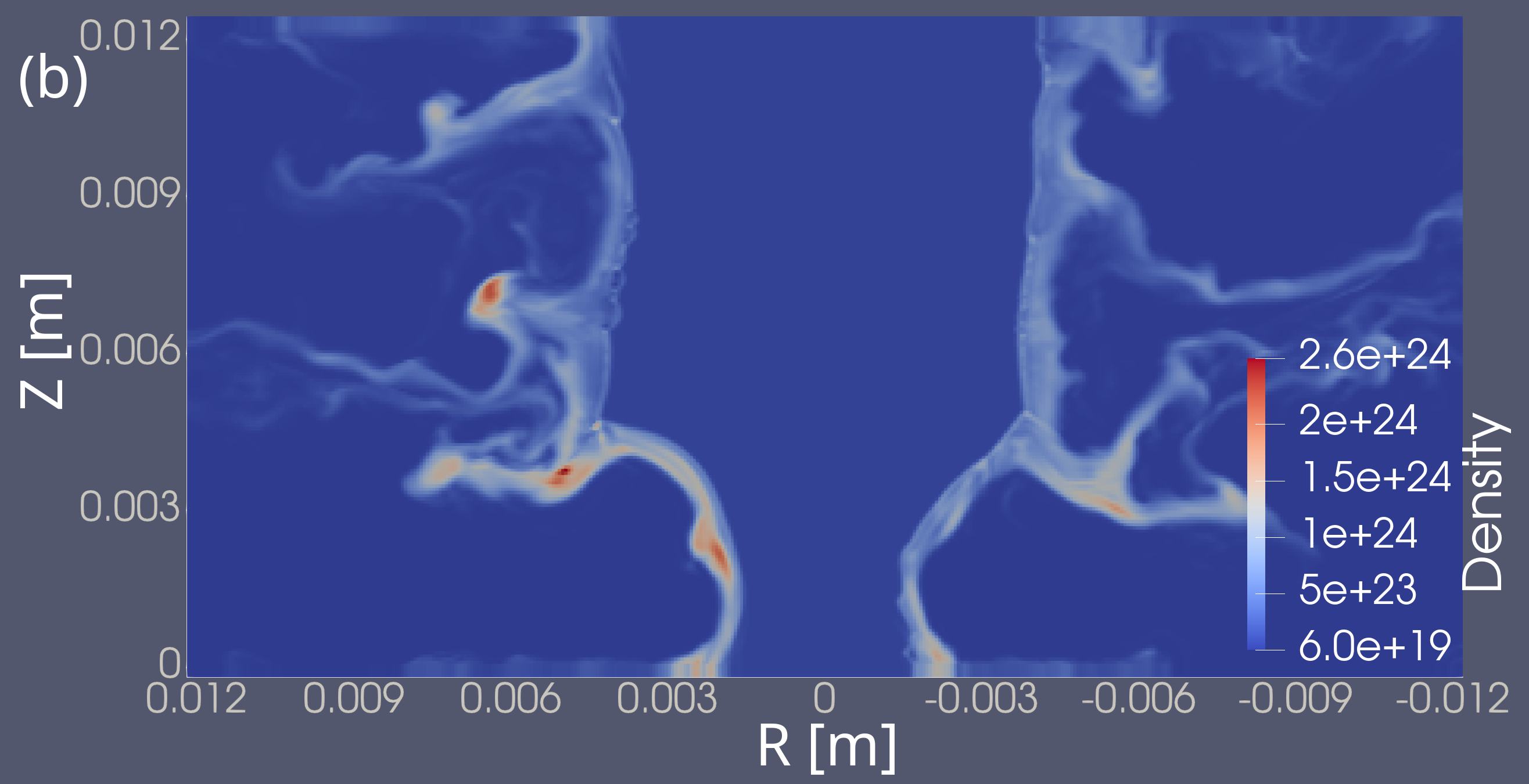}
    \end{tabular}
    \begin{tabular}{cc}
        \includegraphics[width=0.5\linewidth]{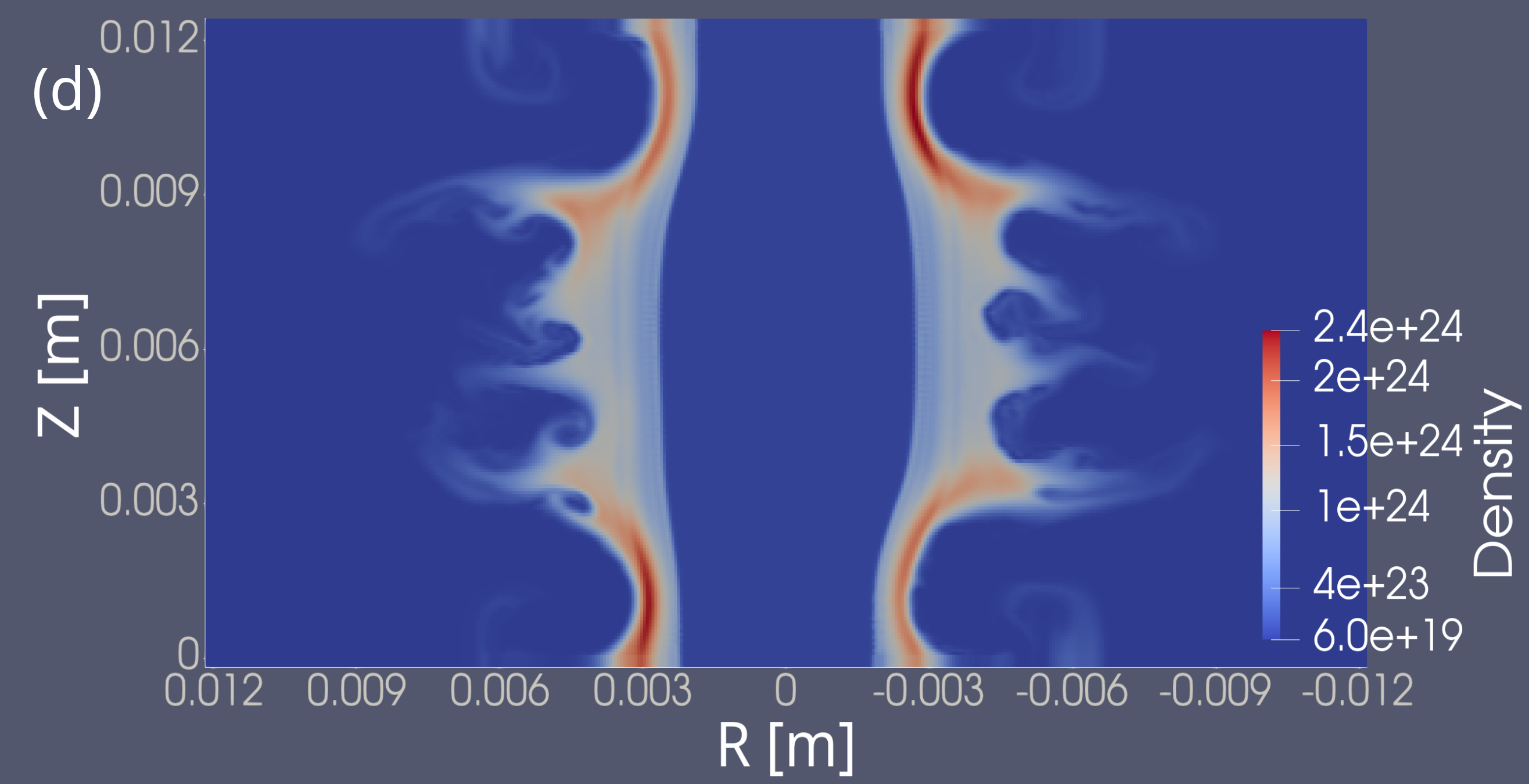}    
        \includegraphics[width=0.5\linewidth]{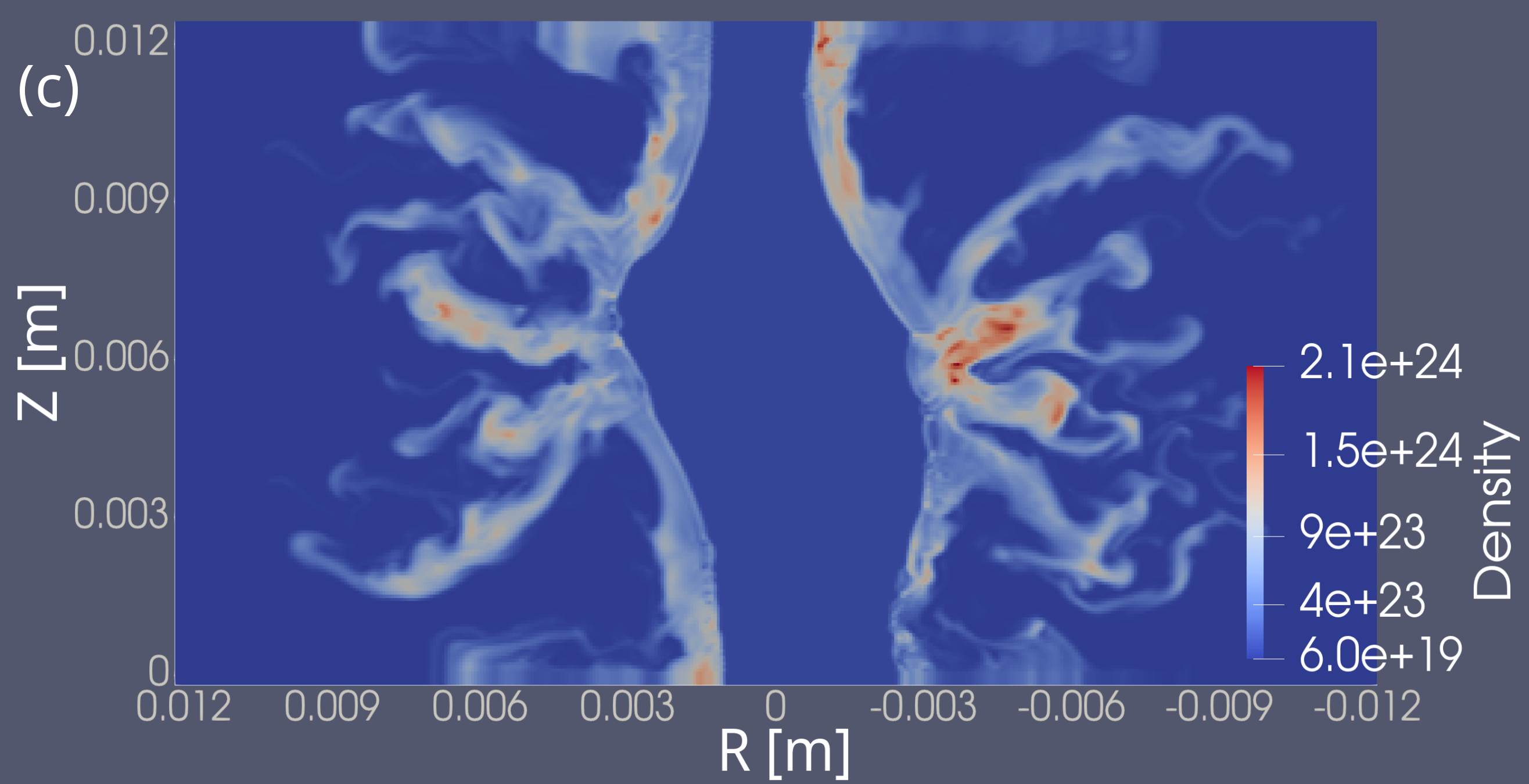}
    \end{tabular}
    \caption{The PERSEUS simulated density profile cut at the $x=0$ plane, where clockwise starting top left (a) - is the Hall MHD model with anomalous resistivity; (b) - the Hall MHD model with Spitzer resistivity; (c) - the pure MHD model with Spitzer resistivity; (d) - the pure MHD model with anomalous resistivity.}
    \label{fig:pers_test}
\end{figure}
\begin{figure*}[tbp]
    \centering
    \begin{tabular}{ccc}
         \includegraphics[width=.3\textwidth]{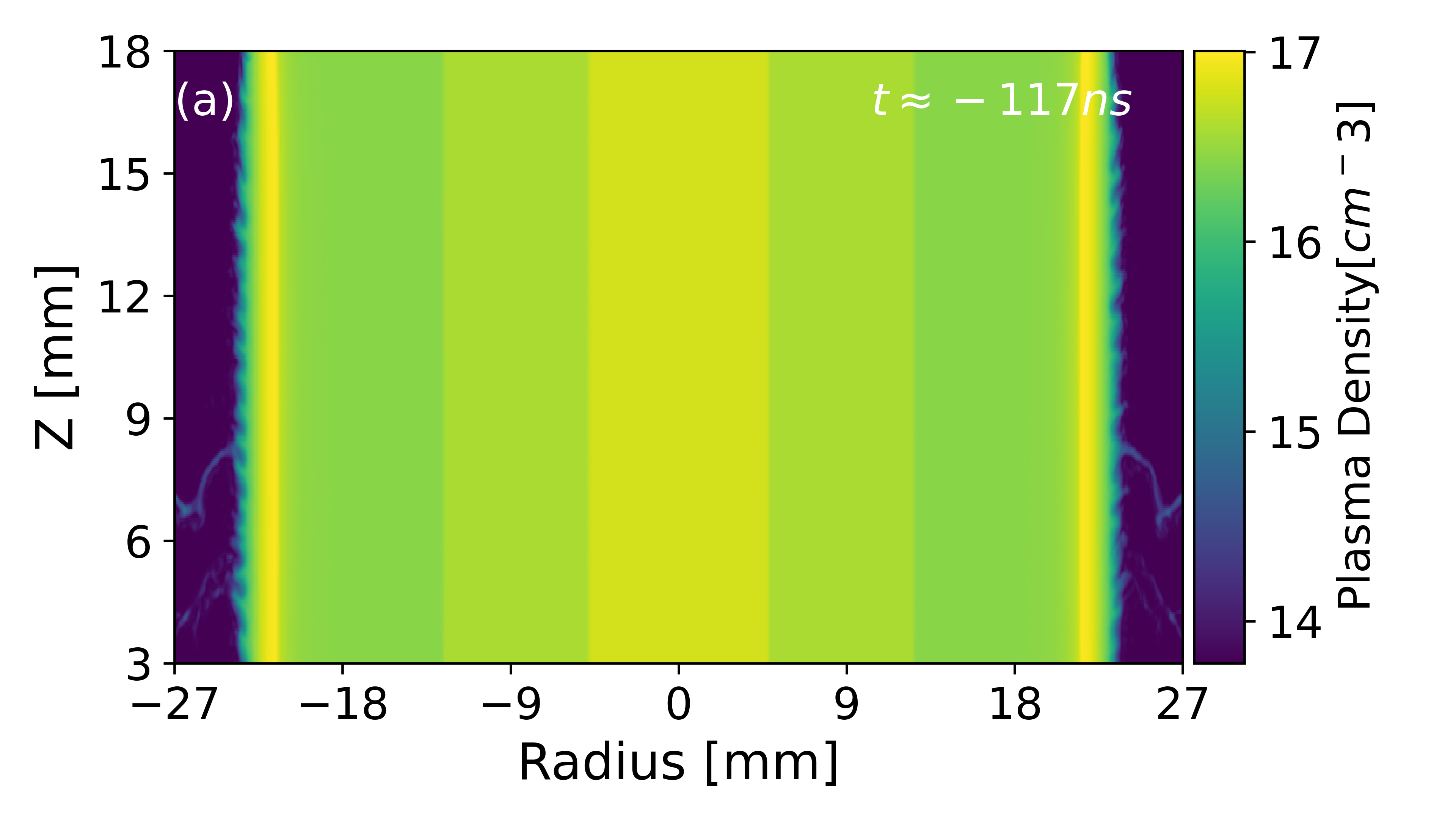}
         \includegraphics[width=.3\textwidth]{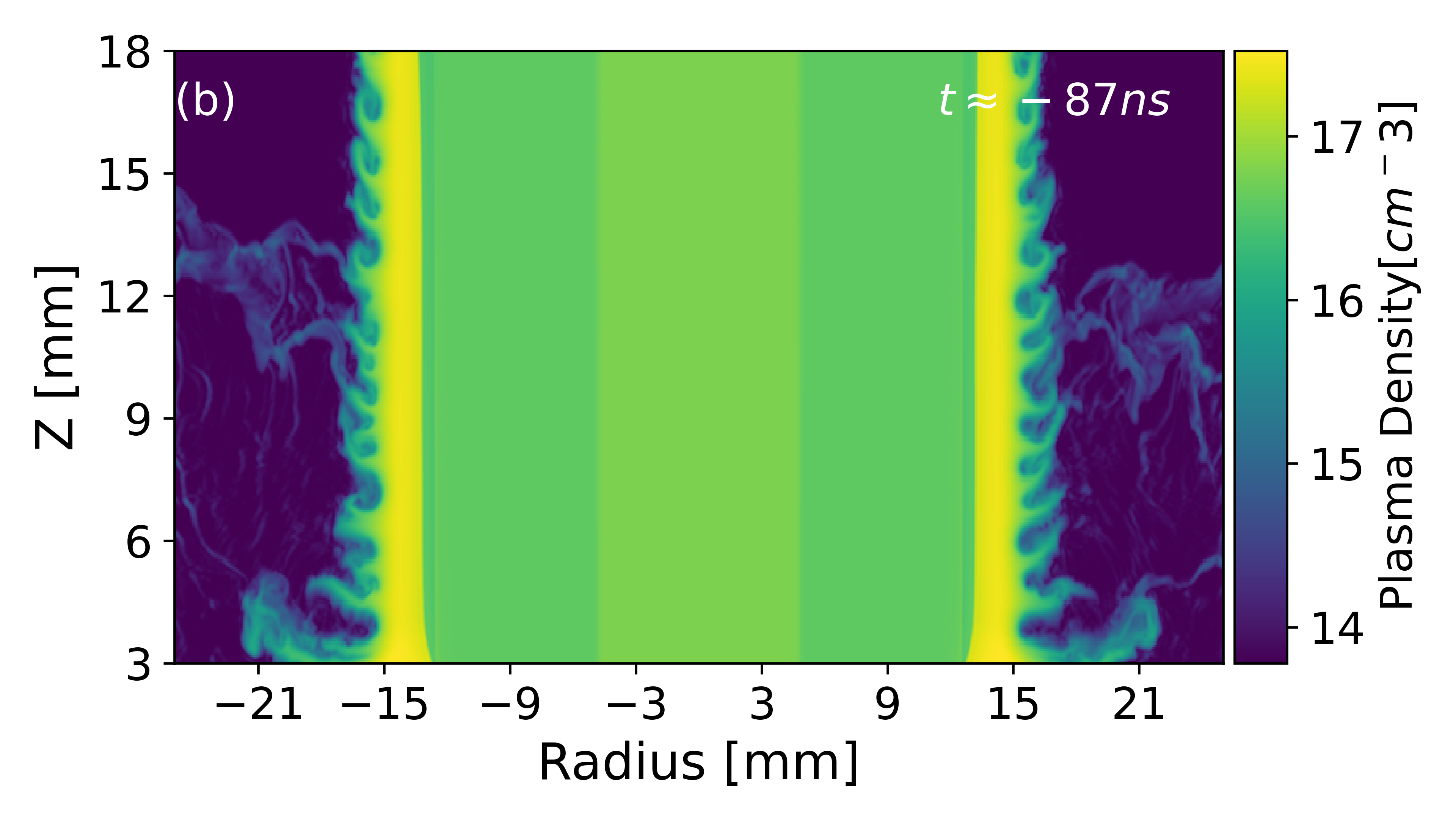}
         \includegraphics[width=.3\textwidth]{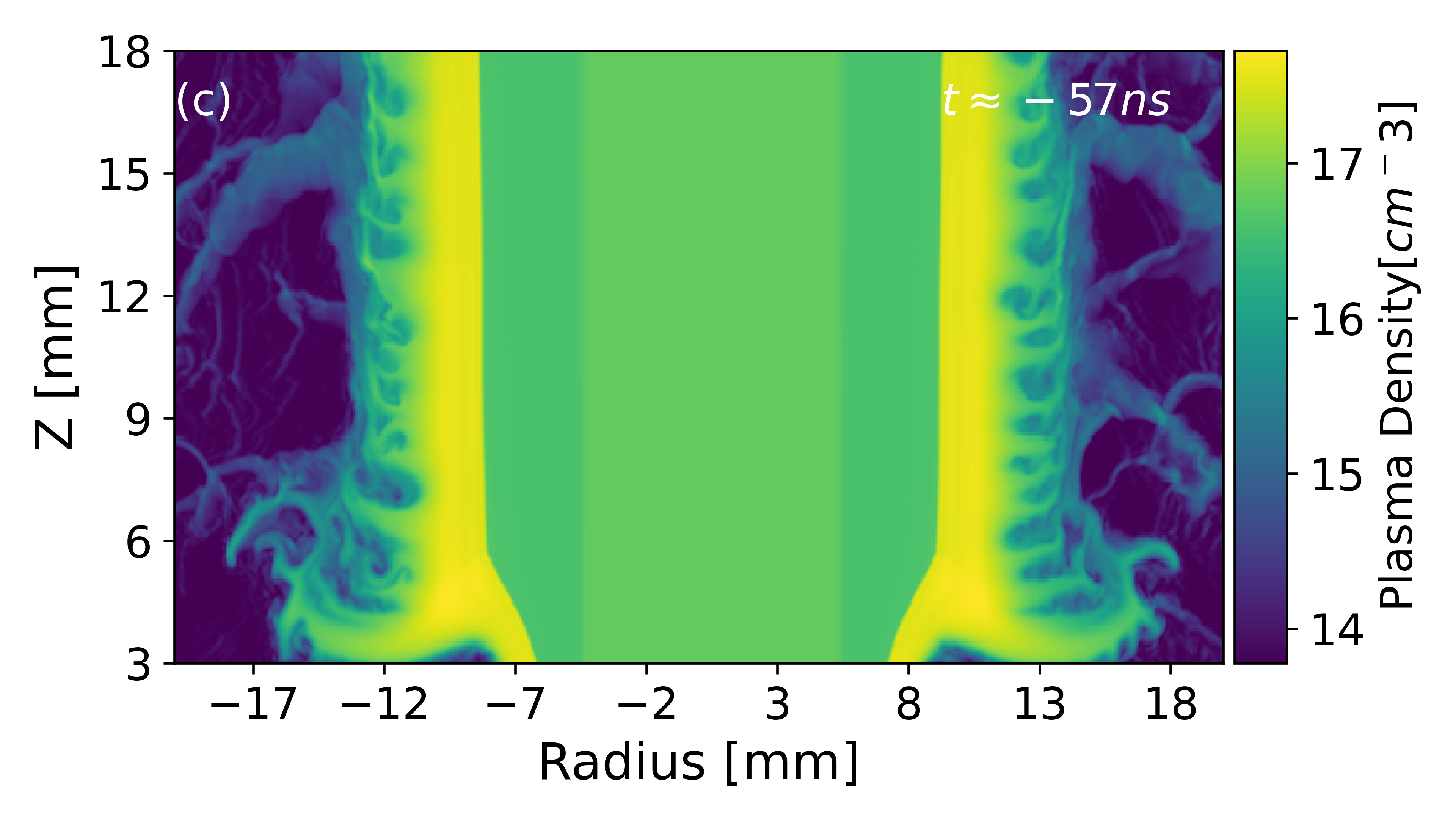}
    \end{tabular}
    \vspace{0.5cm}
    \begin{tabular}{ccc}
         \includegraphics[width=.3\textwidth]{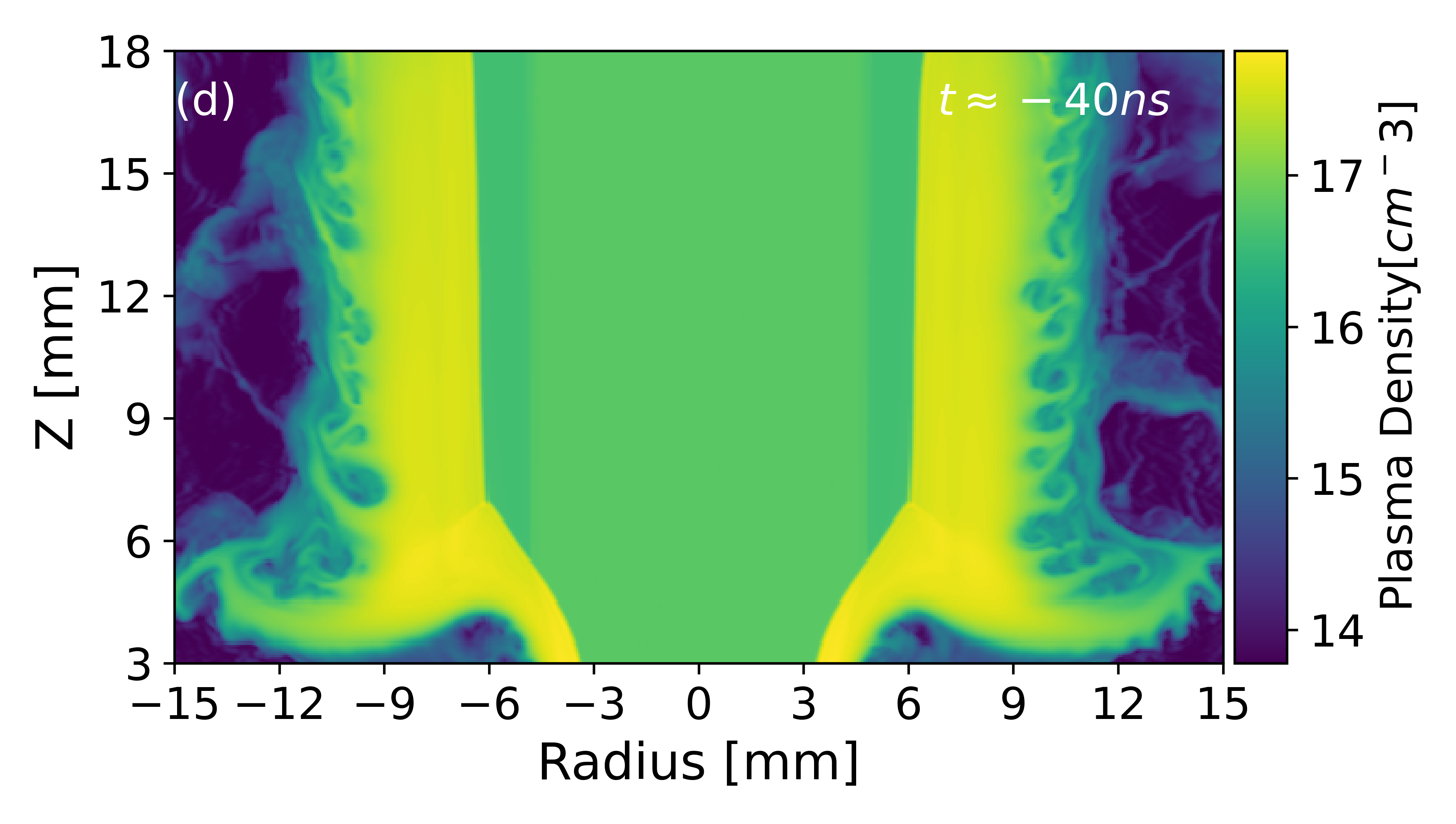}
         \includegraphics[width=.3\textwidth]{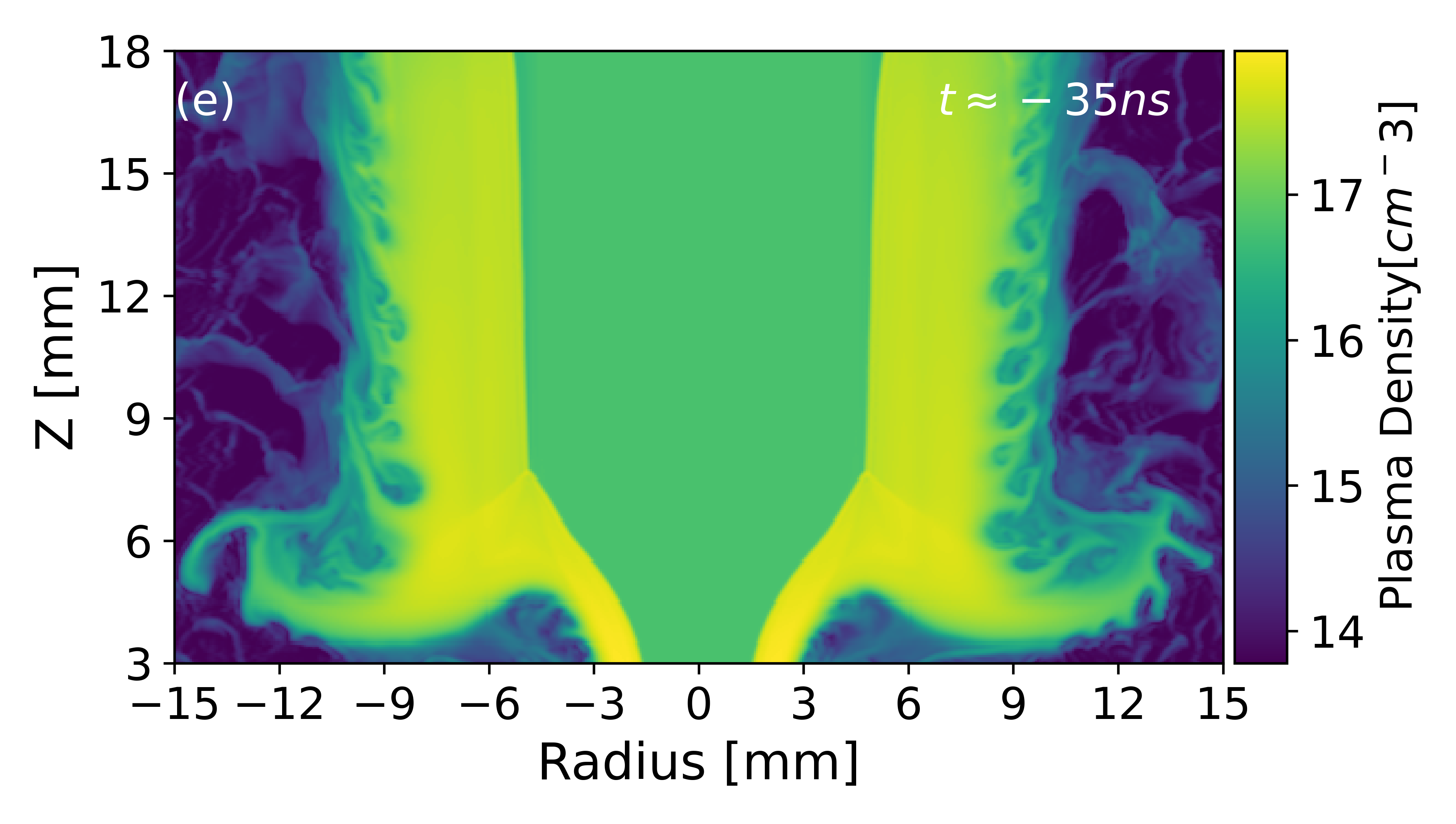}
         \includegraphics[width=.3\textwidth]{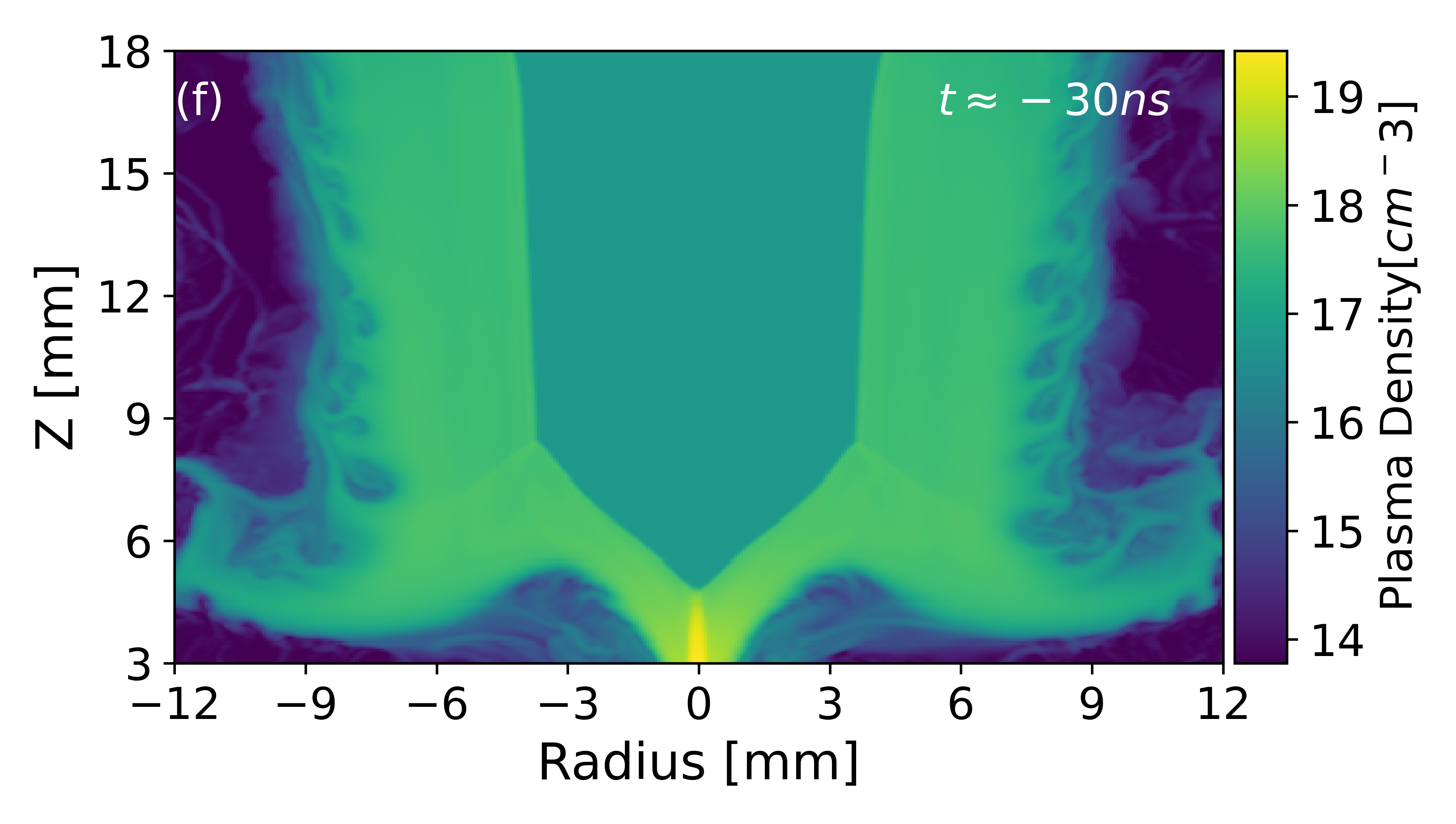}
    \end{tabular}
    \caption{The plasma number density distribution cross-section through the plasma column center on a log scale obtained from the same PERSEUS code run. Timings in the top right of each figure indicate the time relative to the stagnation stage onset.}
    \label{fig:pers_dens}
\end{figure*}
\par We display results from four PERSEUS simulations from the combinations of when the Hall term is turned on and off and when the anomalous resistivity is turned on and off. We run simulations on the Bridges-2 facility in configuration \#1 mentioned above. In Fig. \ref{fig:pers_test} we present the PERSEUS simulated density profiles to highlight the main differences and key features related to the models in question. In Fig. \ref{fig:pers_test}a, we show the Hall MHD with anomalous resistivity simulation run. We note not only that $\mathbf{E}\times\mathbf{B}$ drifts in the MRTI bubbles have the directionality matching experiment, but also early pinching close to the cathode is well described following roughly the size and growth rate measured during the experiment. Moreover, the MRTI spatial wavelength is roughly $2.5mm$; however, although showing a somewhat close value to the experiment (~$3.5~mm$ from Ref. [\onlinecite{Sander_impldyn}]), it is affected by the adjusted plasma column size for this simulation. The other measure is the width of the plasma sheath, as measured from the shock front to the trailing edge of the layer carrying most of the current. The estimated plasma sheath width, measured from the axial current density profile, changes from $<2~mm$ early into the run-in stage, to $\approx3.5~mm$ closer to the stagnation stage onset. That correlates well with $3~mm$ width measured experimentally using Thomson scattering in over-massed Ne implosions.\cite{Sander_impldyn} As seen in Figs. \ref{fig:pers_test}(c,d), it is clear that the pure MHD model with and without anomalous resistivity does not have the right directionality of the $\mathbf{E}\times\mathbf{B}$ drifts while also failing to reproduce asymmetrical early pinch near the cathode surface. Additionally, the plasma sheath width estimates, when the Spitzer resistivity has been used, yield $<0.5~mm$ for both Hall MHD and pure MHD runs. We conclude then, that the dynamics of the MRTI development is governed primarily by the Hall MHD effects while the plasma sheath structure is mainly dependent upon the chosen resistivity model. Therefore, the LHDI-driven anomalous resistivity model represents a better fit for the GZP plasmas.

\par The next step is to analyze the simulation run where the experimental conditions were reproduced as close as possible, i.e. configuration \#2. Due to limited amount of space available on the hard drives, the output included the current density, electric field, magnetic field, velocity in a form of vector fields, and density as a scalar field. Comparison with the experimental measurement, such as laser interferometry and Thomson scattering, is an important step to validate the numerical simulation result. Here, we will use the results from Ref. [\onlinecite{Sander_impldyn}] where these diagnostics were used on Ne over-massed implosions on COBRA that were also the basis for PERSEUS runs. From that paper, we can use the electron/ion temperature and radial velocity profiles measured by Thomson scattering and the Abel inverted local electron density from small shift shearing interferometry. From shot \#5867, the plasma column diameter, measured from the plasma sheath leading edge, was about $36~mm$ at $t=167~ns$, the electron density was ~$1.2\times10^{24}~m^{-3}$ and the sheath width between $\approx1.1~mm$ to $\approx1.4~mm$. The PERSEUS code run in configuration \#2 had only the bulk density as an output, which at this plasma column size equals to $n\approx2\times10^{23}~m^{-3}$, which roughly corresponds, given that $T\approx100~eV$ and the average ionization state $Z=5$, to $\approx10^{24}~m^{-3}$ electron density. The plasma sheath width in the simulation at this plasma column size was $\approx1.6\pm0.13~mm$, which is reasonably close to the experimentally measured value. At a later time, $t\approx210~ns$, the Thomson scattering, based on the same shot, shows the plasma sheath width of about $4~mm$, which was influenced by the MRTI bubble, which likely led to an overestimation. The radial velocity profile of the sheath shows a drop from $\approx150~km/s$ at the leading edge to $\approx100~km/s$ at the trailing edge while the plasma column size is ~$22~mm$. In the simulation, the plasma column size of ~$22~mm$ corresponds with $\approx3~mm$ sheath width. The radial bulk velocity profile has ~$200\pm10~km/s$ and ~$170\pm10~km/s$ at the leading and trailing edges, respectively. To conclude, these comparisons show satisfactory agreement between the Ne gas-puff z-pinch implosion on COBRA and the corresponding PERSEUS simulation.
\begin{figure}
    \centering
    \includegraphics[width=1\columnwidth]{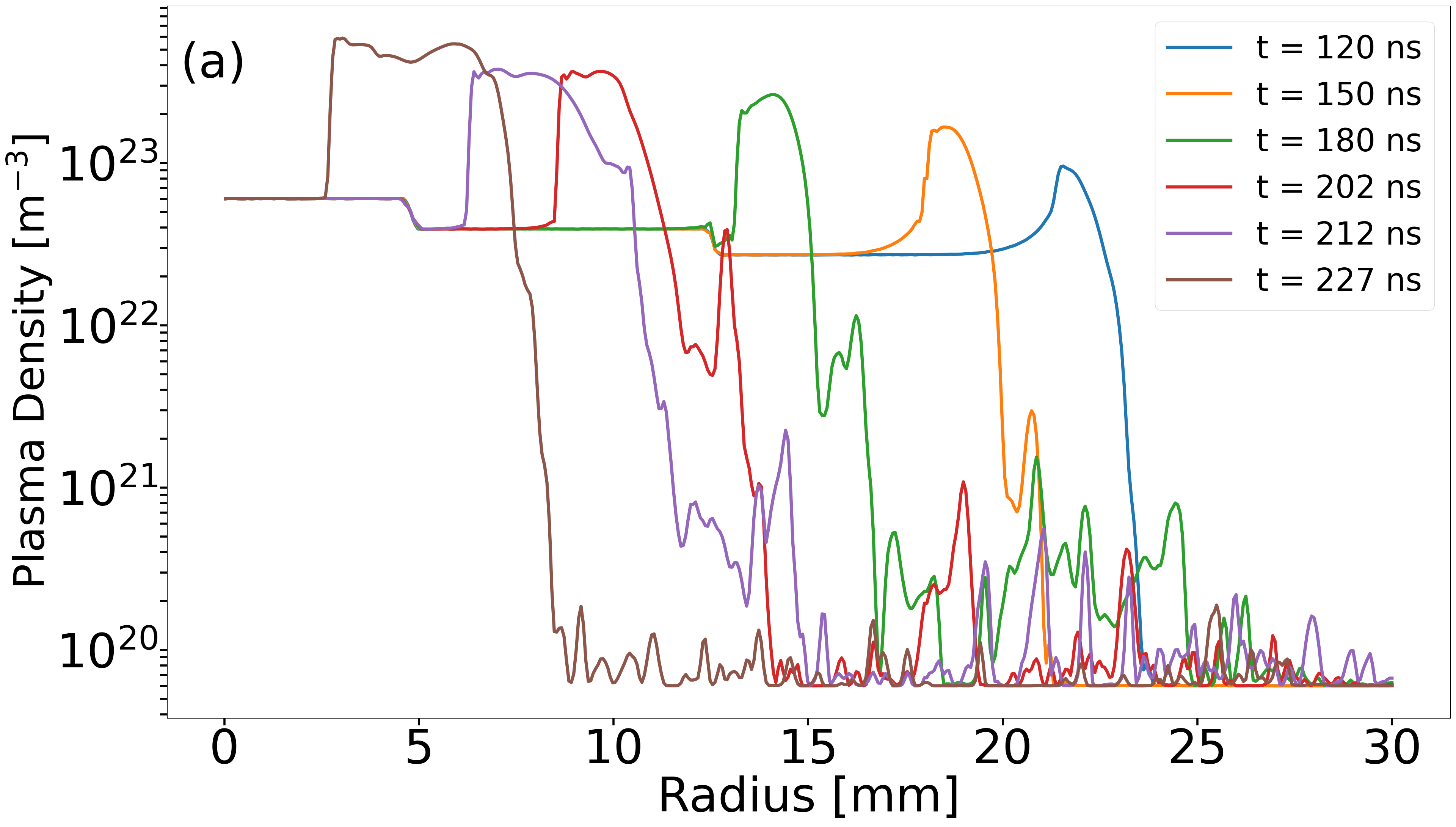}
    \includegraphics[width=1\columnwidth]{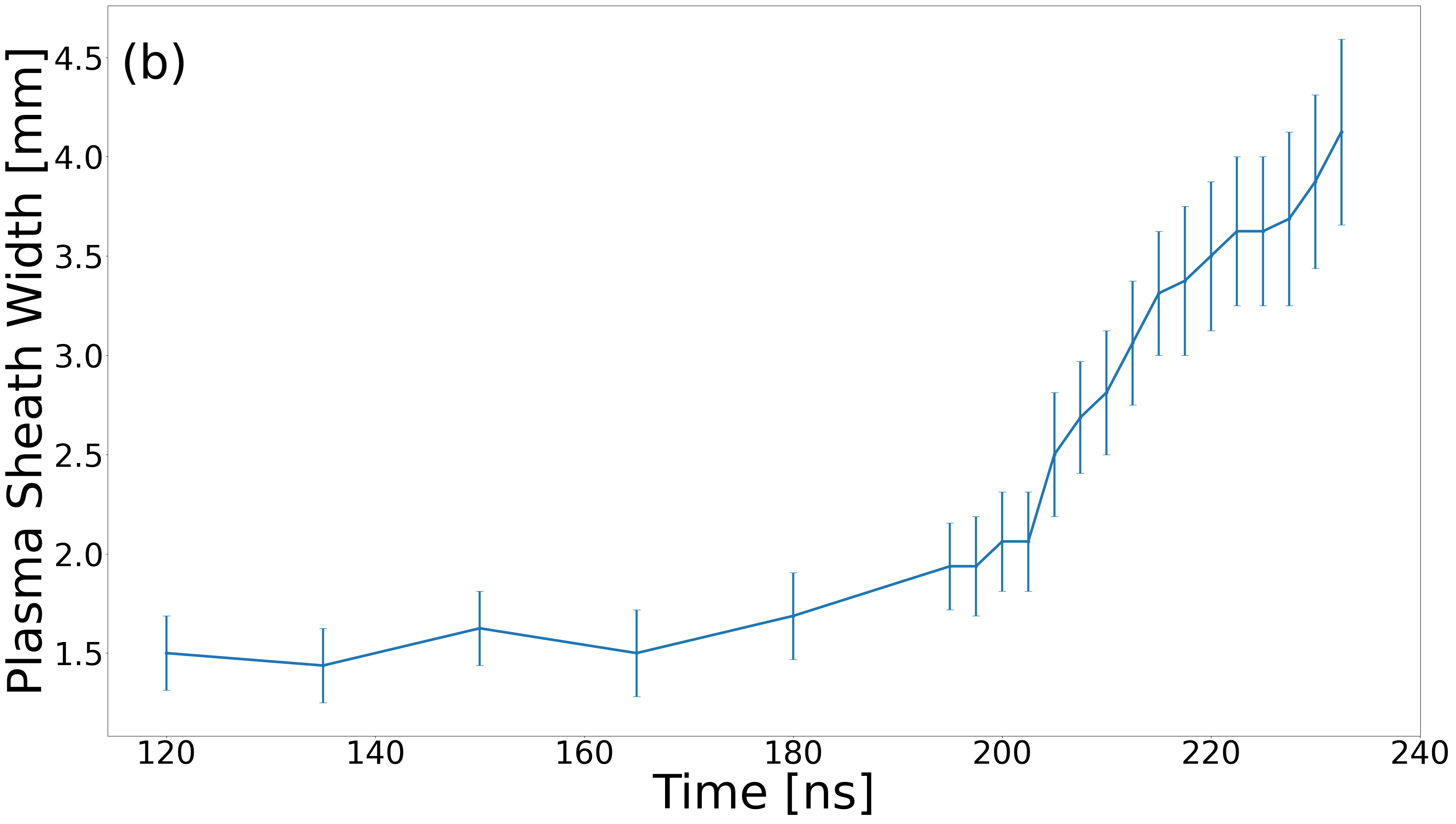}
    \caption{The plasma sheath width as a function of time.}
    \label{fig:wd_evo}
\end{figure}
\par To present the simulated evolution of the plasma column, we show in Fig. \ref{fig:pers_dens} the bulk plasma density profile cross section in the $r-z$ plane of cylindrical coordinate system. In these figures, the main features noted above are clearly apparent, i.e. the MRTI's orientation and spatial frequency and early pinching near the cathode surface. The plasma sheath size evolution is also observed from these plots. To better display sheath dynamics, we show a radial profile of the plasma density at different timesteps in Fig. \ref{fig:wd_evo}(a). To emphasize this evolution further, we show the plasma sheath width versus time in Fig. \ref{fig:wd_evo}(b). To construct this figure and compute the plasma sheath width, we look at the one-dimensional radial plasma density profile, $n(r)$ shown in Fig. \ref{fig:wd_evo}(a), at the pinch column midplane where it roughly has a square function shape and finding its margins (leading and trailing edges) yields the plasma sheath width.

\par To summarize, in this paper we have explored the importance of Hall physics and anomalous resistivity under GZP conditions in the PERSEUS XMHD code. We established, that the Hall term is essential to the morphology of GZP implosions including the vacuum-plasma interfaces and electron drift. The PERSEUS code with incorporated anomalous resistivity was found to qualitatively reproduce the experimentally observed plasma density within the plasma sheath and the evolution of the width. Future work will include the analysis of PERSEUS simulation output, where we observed a radially filamented structures in the parallel current density of currently unknown origin, developing behind the shock front. Furthermore, the experiments are under development to have measurements of the LHDI evolution during GZP implosion.

\par While the results of this study are clearly applicable to gas-puff z-pinches in which the density is sufficiently low for the Hall effect and anomalous resistivity to significantly alter the dynamics compared to resistive MHD, there are other experiments in which these effects could also play an important role. For example, in the study by Seyler et. al,\cite{Seyler_2018} it was found that the Hall effect was important in the dynamics of the compression of low-density plasma in the region surrounding the MagLIF Beryllium liner. The compression of the applied axial magnetic field was found to lead to a helical compression of the liner. While not studied in that paper, it is also possible that the extreme current density in the low-density region was sufficient to involve the LHDI instability. The study of the LHDI in this case should consider the potential stabilizing effect of the strong magnetic shear.\cite{Krall_1977}

\par One of the most important take-aways from the results demonstrated here is that when undertaking of numerical modeling of HEDP experiments, one should carefully consider the resistivity model, and in particular the potential impact of strong currents in low-density regions on the resistivity. Since the LHDI resistivity model we have used is phenomenological and only partially justified, there is a clear need for a more complete and rigorous study of the impact of current-driven instabilities on the resistivity. Although not fully validated, the results presented here make a strong case for the presence of anomalous resistive effects and the inadequacy of standard Spitzer resistivity under the conditions associated with GPZs. Even the advanced resistivity models such as Lee-More-Desjarlais that apply mainly to high-density plasma\cite{Lee_More_Desj} are likely to be inadequate in high-current low-density regions. Although the plasma is low density this does not mean that its effect on the much higher density regions is unimportant.\cite{Seyler_2018,Seyler_2020} This potentially means that Hall and anomalous resistivity effects need to be considered in many HEDP experiments.

\begin{acknowledgments}
This research was supported by the Cornell Laboratory of Plasma Studies, by the Engineering Dean's office through the College Research Incentive Program, by the K. Bingham Cady Memorial Fund, and by the Air Force Office of Scientific Research under award number FA9550-24-1-0066. This work used Bridges-2 at Pittsburgh Supercomputing Center through allocation PHY230055 from the Advanced Cyberinfrastructure Coordination Ecosystem: Services \& Support (ACCESS) program, which is supported by National Science Foundation grants \#2138259, \#2138286, \#2138307, \#2137603, and \#2138296.
\end{acknowledgments}

\section*{Data Availability Statement}
The data that support the findings of this study are available from the corresponding author upon reasonable request.

\bibliography{biblo.bib}

\end{document}